\documentclass[reprint,twocolumn,superscriptaddress,showkeys,
nofootinbib,notitlepage,amsmath,amssymb,floatfix]{revtex4-1}

\pdfoutput=1 \usepackage{latexsym,amsmath,amssymb,lmodern,float,url}
\usepackage{natbib} \usepackage{color} \usepackage{multirow}
\usepackage{bm} \usepackage{array}

\def\de{\delta^{\vphantom{1}}} \def\bde{{\bar\delta}}

\def\h3{{\displaystyle{\frac 3 2}}}

\begin{document}

\title{The Simple Spectrum of $c\bar c c\bar c$
States in the Dynamical Diquark Model} \author{Jesse F. Giron}
\email{jfgiron@asu.edu} \author{Richard F. Lebed}
\email{Richard.Lebed@asu.edu} \affiliation{Department of Physics,
Arizona State University, Tempe, AZ 85287, USA}

\date{August, 2020}

\begin{abstract}
We develop the spectroscopy of $c\bar c c\bar c$ and other all-heavy
tetraquark states in the dynamical diquark model. In the most minimal
form of the model ({\it e.g.}, each diquark appears only in the
color-triplet combination; the non-orbital spin couplings connect
only quarks within each diquark), the spectroscopy is extremely
simple.  Namely, the $S$-wave multiplets contain precisely 3
degenerate states ($0^{++}$, $1^{+-}$, $2^{++}$) and the 7 $P$-wave
states satisfy an equal-spacing rule when the tensor coupling is
negligible.  When comparing numerically to the recent LHCb results,
we find the best interpretation is assigning $X(6900)$ to the $2S$
multiplet, while a lower state suggested at about $6740$~MeV fits
well with the members of the $1P$ multiplet.  We also predict the
location of other multiplets ($1S$, $1D$, {\it etc.}) and discuss the
significance of the $cc$ open-flavor threshold.
\end{abstract}

\keywords{Exotic hadrons, diquarks} 
\maketitle 
\section{Introduction}\label{sec:Intro} 

The LHCb Collaboration has recently presented
evidence~\cite{Aaij:2020fnh} for the observation of at least one
resonance in the $J/\psi$-pair spectrum at about 6900~MeV, and the
likely presence of at least one additional resonance lying below this
mass but above the 6200~MeV $J/\psi$-pair threshold.  Such states are
naturally assigned the valence-quark content $c\bar c c\bar c$,
making them the first all-heavy multiquark exotic candidates claimed
to date in the experimental literature.

Theoretical studies of $c\bar c c\bar c$ states have a much
longer history, dating indeed to a time only two years after the
discovery of the $J/\psi$~\cite{Iwasaki:1976cn} and followed by a
smattering of papers in the 1980s~\cite{Chao:1980dv,Ader:1981db,
Badalian:1985es}.  The current interest in $c\bar c c\bar c$ states,
starting in 2011~\cite{Berezhnoy:2011xy,Berezhnoy:2011xn} and
particularly ramping up since 2016~\cite{Wu:2016vtq,Chen:2016jxd,
Karliner:2016zzc,Wang:2017jtz,Bicudo:2017usw,Debastiani:2017msn,
Anwar:2017toa,Wang:2018poa,Liu:2019zuc,Wang:2019rdo,Bedolla:2019zwg,
Chen:2020lgj,Liu:2020eha,Wang:2020ols,Jin:2020jfc,Yang:2020rih,
Becchi:2020uvq,Lu:2020cns,Chen:2020xwe,Wang:2020gmd}, emerged from the
expectation of dedicated searches at the LHC\@.

A notable feature of the all-heavy multiquark exotics $Q_1 \overline
Q_2 Q_3  \overline Q_4$ ($Q_i \! = \! c$ or $b$), in contrast to the
known exotics $Q \overline Q q \bar q^\prime$~\cite{Lebed:2016hpi}
($q, q^\prime \! \in \! \{ u,d \}$), is the lack of a plausible
molecular structure for the states.  The lightness of the quarks
$q, \bar q^\prime$ in the $Q \overline Q q \bar q^\prime$ case
suggests the possibility of $(Q \bar q^\prime) (\overline Q q)$
molecules, bound by the exchange of light mesons with valence content
$(q \bar q^\prime)$ and possessing a spatial extent at least as large
as the light-meson wave function, of order $1/\Lambda_{\rm QCD} \!
\simeq \! O(1)$~fm.  If the state lies especially close to the
$(Q \bar q^\prime)(\overline Q q)$ threshold [{\it e.g.}, $X(3872)$],
then its spatial extent is determined by the inverse of the binding
energy and can be quite substantial, possibly as large as several fm.
Moreover, Yukawa-like light-meson binding exchanges as an explanation
for such near-threshold states begin to appear implausibly
fine-tuned, and instead {\it threshold rescattering effects\/} (loop
exchanges of virtual particles between the constituent mesons that
numerically enhance the amplitude near the threshold) provide a
mechanism for binding the state.  In contrast, the case of all-heavy
$Q_1 \overline Q_2 Q_3 \overline Q_4$ states lacks a light-meson
exchange mechanism, both for Yukawa-type exchanges and for threshold
effects.  The $X(6900)$ is noted~\cite{Aaij:2020fnh} to lie in the
vicinity of the $\chi_{c0} \chi_{c0}$ and $\chi_{c1} \chi_{c0}$
thresholds, but to our knowledge no calculation has yet suggested the
ability of such a threshold rescattering to produce a strong
resonance.

In general, one expects the lowest-lying $Q_1 \overline Q_2 Q_3
\overline Q_4$ states to exhibit comparable distances between all
four heavy quarks.  If, say, the $Q_1 \overline Q_2$ and
$Q_3 \overline Q_4$ pairs are formed with substantially smaller
internal separations than the distance between the two pairs, then
one expects the immediate formation of two free conventional
quarkonium states rather than a single resonance, even if both pairs
are in color octets and require gluon exchange (which has a range
comparable to that of light-meson exchange) in order for $Q_1
\overline Q_2$ and $Q_3 \overline Q_4$to hadronize as color singlets.

As a result, the most common models for $Q_1 \overline Q_2 Q_3
\overline Q_4$ states assume a diquark-antidiquark
[($Q_1 Q_3$)($\overline Q_2 \overline Q_4$)] structure, typically
exploiting the attractive color-antitriplet quark-quark coupling.
One should keep in mind, however, that if all four quarks have
comparable separations (as is anticipated for the ground states),
then a combination of different color structures should be expected
to appear for those states ({\it e.g.}, as in the lattice simulation
of Ref.~\cite{Bicudo:2017usw}).

Beyond the ground states, the separations between the quarks can
become differentiated.  As noted above, closer association of the
$Q \overline Q$ pairs is expected to lead to an immediate
dissociation into quarkonium pairs, while the configuration
($Q_1 Q_3$)($\overline Q_2 \overline Q_4$) with color-triplet
diquarks becomes the only one that features an attractive interaction
between the component constituents (the quarks within the diquarks),
but must still remain bound due to confinement, independent of the
exchange of any number of gluons.  These features define the
{\it dynamical diquark picture\/} of multiquark
exotics~\cite{Brodsky:2014xia,Lebed:2015tna}.  In the original
picture, the diquark separation is a consequence of the production
process; for example, $c\bar c q\bar q^\prime$ tetraquarks can be
manifested due to the large momentum release between the $c\bar c$
pair in $B$-meson decays into a $(cq)(\bar c \bar q^\prime)$
structure.  To be more precise, the diquark-antidiquark state couples
most strongly to the portion of the four-quark momentum-space wave
function for which the relative momentum between the quasiparticles
$\de \! \equiv \! (Q_1 Q_3)$ and $\bde \! \equiv \! (\overline Q_2
\overline Q_4)$ is significantly larger than the relative momenta
within them.

The dynamical diquark picture is elevated to a full model by
identifying its mass eigenstates with those of the gluon field
connecting the diquarks~\cite{Lebed:2017min}.  Explicitly,
confinement limits the eventual separation of the $\de$-$\bde$ pair
even though they may form with a large relative momentum, and the
specific stationary states of the full system are supplied by the
quantized modes of the gluon field stretching between the two heavy,
(eventually) nearly stationary sources $\de,\bde$.  This approach
uses the Born-Oppenheimer (BO) approximation in precisely the same
manner as is done for simulations of heavy-quark hybrids on the
lattice ({\it e.g.}, Refs.~\cite{Juge:1997nc,Juge:1999ie,Juge:2002br,
Morningstar:2019,Capitani:2018rox}).  Indeed, the specific form of
the static potential $V_\Gamma(r)$ between the heavy sources for a
particular BO glue configuration $\Gamma$ is precisely the same one
computed in each lattice simulation just referenced.  The
corresponding coupled Schr\"{o}dinger equations were first
numerically solved for $c\bar c q\bar q^\prime$ states in
Ref.~\cite{Giron:2019bcs}.

Typical diquark models approximate the quasiparticles $\de, \bde$ to
be pointlike, even though they are expected to have spatial extents
comparable to those of mesons carrying the same valence-quark
flavor content.
Nevertheless, model calculations in Ref.~\cite{Giron:2019cfc} for
$c\bar c q\bar q$ states show that finite diquark size has a
surprisingly mild effect on the spectrum for a $\de \! = \! (cq)$
radius as large as 0.4~fm.

The dynamical diquark model also selects a very specific set of
spin-dependent couplings as the ones deemed most physically
significant.  In this model the $\de, \bde$ pair form
distinguishable, separate entities within the full state, so that the
dominant spin-spin couplings are taken to be the ones between quarks
within each diquark~\cite{Maiani:2014aja}, while typical existing
models for $c\bar c c\bar c$ states ({\it e.g.},
Refs.~\cite{Berezhnoy:2011xy,Berezhnoy:2011xn}) treat all quark
spin-spin interactions on equal footing, or consider only couplings
to full diquark spins ({\it e.g.}, Ref.~\cite{Bedolla:2019zwg}).  The
more restrictive paradigm used here leads to very simple predictions
for the spectrum of $c\bar c c\bar c$ states, particularly in
$S$-wave multiplets, which will become immediately testable once the
{quantum numbers of the $c\bar c c\bar c$ states are known.

On the other hand, the dominant operators in this model for
$c\bar c c\bar c$ states carrying orbital angular momentum dependence
(relevant to $P$- and higher-wave states)  are taken to couple only
to the diquarks as units, since $\de,\bde$ are assumed to have no
internal orbital excitation for all low-lying $c\bar c c\bar c$
states.\footnote{In contrast, the tensor operator for $P$-wave
$c\bar c q\bar q^\prime$ states in Ref.~\cite{Giron:2020fvd}, owing
its origin to a pionlike exchange within the state, was chosen to
couple only to the light-quark spins within the diquarks.
Nevertheless, the matrix elements for an alternative tensor operator
that couples only to the full diquark spins (as to be used here) are
also computed in that work.}  The resultant spin-orbit and tensor
operators for the low-lying spectrum are the same as those used in
Ref.~\cite{Bedolla:2019zwg}, but differ from those used in
Ref.~\cite{Liu:2020eha}, which instead are chosen to couple to all
individual quark spins.  Again, a very simple spectrum arises in this
model for the $P$-wave states, the degree of validity for which will
become immediately apparent with further data.

Our purpose in this paper is therefore not to compete with detailed
calculations of spectra that are based upon assuming specific forms
for all operators contributing to the Hamiltonian of
$c\bar c c\bar c$ states ({\it e.g.}, using a one-gluon-exchange
potential to obtain an explicit functional form for the coefficient
for every operator, as in Ref.~\cite{Bedolla:2019zwg}).  Rather, we
describe the most significant features in the spectrum
parametrically, identifying particular spin-spin, spin-orbit, or
tensor terms to pinpoint their origin, while remaining agnostic as to
the precise dynamical origin of these operators.  We nevertheless
also present an initial fit to the $c\bar c c\bar c$ spectrum, using
numerical values for the Hamiltonian parameters obtained from the
analogous operators in other sectors of exotics to which the model
has previously been applied.  Specifically, the strength of the
spin-spin operator is obtained from a recent fit to $c\bar c s\bar s$
candidates~\cite{Giron:2020qpb}, and the spin-orbit and tensor
strengths are taken from a recent fit to $P$-wave $c\bar c q\bar
q^\prime$ candidates~\cite{Giron:2020fvd}.

This paper is organized as follows.  In Sec.~\ref{sec:Spectrum} we
review the spectroscopy of the model for $S$- and $P$-wave
$Q_1 \overline Q_2 Q_3 \overline Q_4$ states, identifying
quantum-number restrictions arising from spin statistics.
Section~\ref{sec:MassOps} presents the Hamiltonian and tabulates all
matrix elements for the allowed states, and we identify features of
the spectrum that appear based upon their parametric analysis.  In
Sec.~\ref{sec:Num} we present a numerical prediction for the
$c\bar c c\bar c$ spectrum, using as described above the results of
previous work; and in Sec.~\ref{sec:Concl} we conclude.

\section{Spectroscopy of $Q\bar Q Q\bar Q$ Exotics}
\label{sec:Spectrum}

The spectroscopy of $\de$-$\bde$ states in which the diquarks
$\de,\bde$ contain no internal orbital angular momentum, but that
allows for arbitrary orbital excitation and gluon-field excitation
between the $\de$-$\bde$ pair, is presented in
Ref.~\cite{Lebed:2017min}.  For the all-heavy states with
distinguishable quarks in $\de$ and $\bde$ ({\it i.e.}, $b\bar b
c\bar c$, or for that matter, $c\bar c s\bar s$), precisely the same
enumeration of states occurs.  The core states, expressed in the
basis of good diquark-spin eigenvalues with labels such as $1_\de$,
are given by
\begin{eqnarray}
J^{PC} = 0^{++}: & \ & X_0 \equiv \left| 0_\de , 0_\bde
\right>_0 \, , \ \ X_0^\prime \equiv \left| 1_\de , 1_\bde \right>_0
\, , \nonumber \\
J^{PC} = 1^{++}: & \ & X_1 \equiv \frac{1}{\sqrt 2} \left(
\left| 1_\de , 0_\bde \right>_1 \! + \left| 0_\de , 1_\bde \right>_1
\right) \, , \nonumber \\
J^{PC} = 1^{+-}: & \ & Z \ \equiv
\frac{1}{\sqrt 2} \left( \left| 1_\de , 0_\bde \right>_1 \! - \left|
0_\de , 1_\bde \right>_1 \right) \, , \nonumber \\ & \ & Z^\prime \,
\equiv \left| 1_\de , 1_\bde \right>_1 \, , \nonumber \\
J^{PC} = 2^{++}: & \ & X_2 \equiv \left| 1_\de , 1_\bde \right>_2 \,
,
\label{eq:Swavediquark}
\end{eqnarray}
with the outer subscripts on the kets indicating total quark spin
$S$.  On their own, these 6 states fill the lowest multiplet
$\Sigma^+_g(1S)$ within the Born-Oppenheimer (BO) approximation for
the gluon-field potential connecting the $\de$-$\bde$ pair.  Higher
BO potentials (like $\Sigma^-_u$, where standard BO quantum-number
labels such as these are defined in Ref.~\cite{Lebed:2017min})
produce the multiquark analogues to hybrid mesons, and thus are
expected to lie about 1~GeV above the $\Sigma^+_g(1S)$ ground states.
For phenomenological reasons to be discussed in Sec.~\ref{sec:Num},
we do not discuss such states further here.

The diquarks $\de,\bde$ in this model transform as color
(anti)triplets, which are antisymmetric under quark-color exchange.
If the quarks within $\de$ or $\bde$ are identical, then the
space-spin wave function of the corresponding diquark must be
symmetric in order to satisfy Fermi statistics for the complete $\de$
or $\bde$ wave function; however, since the model assumes no orbital
excitation within the diquarks, their spatial wave function and hence
also their spin wave function alone must be symmetric, which thus
requires the corresponding diquark spin to equal unity: Only $1_\de$
and $1_\bde$ survive.  In the $c\bar c c\bar c$ or $b\bar b b\bar b$
case, one immediately sees from Eq.~(\ref{eq:Swavediquark}) that the
states $X_0$, $X_1$, and $Z$ are forbidden by spin
statistics.\footnote{One may also consider truly exotic states like
$b\bar b b\bar c$, in which $0_\de$ is
forbidden but $0_\bde$ is allowed, in which case only the state $X_0$
is eliminated.  For such states $C$ also ceases to be a good quantum
number, so that $X_1$ and $Z$ become the same $1^+$ state, thus
leaving a total of 4 states in the multiplet $\Sigma^+_g(1S)$.  In
contrast, the case $b\bar b c\bar c$ (considered in, {\it e.g.},
Ref.~\cite{Bedolla:2019zwg}) retains the $C$ quantum number and all 6
$\Sigma^+_g(1S)$ states.}  The ground-state multiplet
$\Sigma^+_g(1S)$ is thus halved: Only the three states $X^\prime_0$
($0^{++}$), $Z^\prime$ ($1^{+-}$), and $X_2$ ($2^{++}$) survive.  An
identical analysis applies to all radial-excitation multiplets
$\Sigma^+_g(nS)$.

One immediate conclusion of this model becomes evident: If the full
state wave function contains a component that allows either diquark
to appear in the (symmetric) color sextet, then that diquark in the
low-lying states must appear in the antisymmetric spin-0 combination
$0_\de$ or $0_\bde$.  In that case, the full spectrum of 6 states
from Eq.~(\ref{eq:Swavediquark}), most notably a state with
$J^{PC} \! = \! 1^{++}$, must appear.  {\em The observation of a
$1^{++}$ $c\bar c c\bar c$ state in the lowest multiplet (or any
$S$-wave multiplet) would provide direct evidence of dynamics lying
outside the most restrictive diquark models.}

The addition of a nonzero orbital-excitation quantum number $L$ is
now straightforward.  Since the intrinsic parity factor $(-1)$ for
an antiquark appears twice, the parity eigenvalue of the full state
is just given by the usual spatial factor $(-1)^L$.  All $S$-wave, 
$D$-wave, {\it etc.}\ states therefore have $P \! = \! +$, and all
$P$-wave, $F$-wave, {\it etc.}\ states have $P \! = \! -$.  Starting
with the $S$-wave ``core'' states $X^\prime_0$, $Z^\prime$, and $X_2$
of Eqs.~(\ref{eq:Swavediquark}), one invokes the usual angular
momentum addition rules to produce states of good total $J$
(indicated by a superscript ``($J$)'', using the notation developed
in Ref.~\cite{Lebed:2017min}).  Explicitly, the 7 $P$-wave
$c\bar c c\bar c$ states,  accompanied by their $J^{PC}$ eigenvalues,
are
\begin{eqnarray}
& & X^{\prime \, (1)}_{0 \, P} \, (1^{--}) , \ Z^{\prime \, (0)}_P
(0^{-+}) , \ Z^{\prime \, (1)}_P (1^{-+}) , \ Z^{\prime \, (2)}_P
(2^{-+})  , \nonumber \\
& & X^{(1)}_{2 \, P} \, (1^{--}) , \; \ X^{(2)}_{2 \, P} \, (2^{--}) , \
X^{(3)}_{2 \, P} \, (3^{--}) . 
\end{eqnarray}

For completeness, we note that each of the $D$-wave, $F$-wave,
{\it etc.}\ multiplets each contain precisely 9 $c\bar c c\bar c$
states.  In particular, the $\Sigma^+_g(1D)$ multiplet is the lowest
one to contain a $1^{++}$ state, $X_{2 \, D}^{(1)}$.

\section{Mass Hamiltonian} \label{sec:MassOps}

The full mass spectrum of all states in the dynamical diquark model
is computed by the following procedure: First, a particular BO
potential $\Gamma$ ($= \! \Sigma^+_g$, $\Pi_u$, {\it etc.})
that gives rise to a multiplet of states [$\Sigma^+_g(1P)$,
$\Pi_u(2P)$, {\it etc.}] is specified.  The corresponding potentials
$V_\Gamma (r)$ have been computed numerically on the
lattice~\cite{Juge:1997nc,Juge:1999ie,Juge:2002br,Morningstar:2019,
Capitani:2018rox}.  One specifies a diquark mass $m_{\de,\bde}$ (or
in the case of pentaquarks, a color-triplet {\em triquark\/} mass as
well), and solves the resulting Schr\"{o}dinger equation for this
Hamiltonian $H_0$ numerically~\cite{Giron:2019bcs},\footnote{In some
cases the BO potentials mix, leading to coupled Schr\"{o}dinger
equations that require a more involved numerical solution technique.}
giving rise to a multiplet-average mass eigenvalue $M_0(nL)$ for
particular radial ($n$) and orbital ($L$) quantum numbers attached to
the particular BO potential $\Gamma$.  In this paper we are
interested only in the $\Sigma^+_g$ potential, and primarily in the
levels within the lowest multiplets $\Sigma^+_g(1S)$,
$\Sigma^+_g(1P)$, and $\Sigma^+_g(2S)$.

The next step is to identify and compute fine-structure corrections
to the spectrum of each such multiplet.  In the dynamical diquark
model the dominant spin-dependent, isospin-independent operator is
taken to be the spin-spin coupling between quarks in the diquark, and
between the antiquarks in the antidiquark.  In the case of
$Q\overline Q q\bar q^\prime$ states (where $q, q^\prime \! \in \!
\{ u,d \}$) the model also includes a spin-dependent,
isospin-dependent operator that mimics the form present in pion
exchange.  The analysis of the $\Sigma^+_g(1S)$ multiplet of
$c\overline c q\bar q^\prime$ states in Ref.~\cite{Giron:2019cfc}
uses a Hamiltonian consisting only of $H_0$ and the 2 operators thus
described:
\begin{eqnarray}
H & = &
H_0 + 2 \kappa_{qQ} ({\bf s}_q \! \cdot \! {\bf s}_Q + {\bf s}_{\bar
q^\prime} \! \cdot \! {\bf s}_{\bar Q}) + V_0 \, {\bm \tau}_q \!
\cdot \! {\bm \tau}_{\bar q^\prime} \; {\bm \sigma}_q \! \cdot \!
{\bm \sigma}_{\bar q^\prime} \, , \hspace{1em}
\label{eq:Ham}
\end{eqnarray}
where of course $Q \! = \! c$, and $\kappa_{qQ}$ is assumed to be
isospin-symmetric.  This very simple Hamiltonian is used to great
effect in Ref.~\cite{Giron:2019cfc}, where it provides a natural
explanation for the $1^{++}$ $X(3872)$ being the lightest observed
state in the $\Sigma^+_g(1S)$ multiplet and for the appearance of the
preferential decay patterns $Z_c(3900) \! \to \! J/\psi$ and
$Z_c(4020) \! \to \! h_c$.  In the intermediate case of
$c\bar c s\bar s$ states in Ref.~\cite{Giron:2020qpb} as well as in
the all-heavy case $Q\overline Q Q\overline Q$ considered here (or
more generally, $Q_1 \overline Q_2 Q_3 \overline Q_4$), the
isospin-dependent term $V_0$ is absent.  In addition, the
coefficients $\kappa_{qQ}$, $\kappa_{sQ}$, and $\kappa_{QQ}$ refer to
spin couplings within diquarks containing increasingly heavy quarks,
and therefore the diquarks are expected to be increasingly spatially
compact.  Since the fundamental quark spins thus interact at
increasingly close range, one may expect the numerical size of these
couplings to increase for heavier quark combinations, a point to
which we return in Sec.~\ref{sec:Num}.

The $S$-wave Hamiltonian for $Q \overline Q Q \overline Q$
therefore contains only one new parameter,
\begin{equation}
H =
H_0 + 2 \kappa_{QQ} ({\bf s}_Q \! \cdot \! {\bf s}_Q + {\bf s}_{\bar
Q} \! \cdot \! {\bf s}_{\bar Q}) \, ,
\label{eq:Swavecccc}
\end{equation}
where the two factors of ${\bf s}_Q$ and of ${\bf s}_{\bar Q}$ are
each understood to apply to a separate heavy quark.  The eigenvalues
of $H$ are trivially computed in the basis of good diquark spin:
\begin{equation}
M  = M_0 + \kappa_{QQ} \left[ s_\de (s_\de + 1) + s_{\bde}
(s_{\bde} + 1) - 3 \right] \, .
\end{equation}
Since as noted above, $s_\de \!  = \! s_\bde \! = \! 1$ in any state
for which diquarks have negligible coupling to the color-sextet
channel, we immediately obtain a strong result: {\em The 3 states of
each $\Sigma^+_g(nS)$ multiplet, $0^{++}$, $1^{+-}$, and $2^{++}$,
are degenerate in this model, with a common mass eigenvalue given by}
\begin{equation}
M(nS) = M_0 + \kappa_{QQ} \, ,
\label{eq:SwaveMass}
\end{equation}
where of course both $M_0$ and $\kappa_{QQ}$ may vary with the radial
excitation number $n$.  The measurement of nonzero mass splittings
between these three states would therefore provide direct evidence
that the quarks within different diquarks have nonnegligible
spin-spin couplings between them.\footnote{This result is
parametrically apparent from the first equations of Sec.~IIB in
Ref.~\cite{Berezhnoy:2011xn} (setting their $\kappa_+ \!  = \! 0$).
However, since all spin-spin couplings are numerically comparable in
their model, this feature was not commented upon there.}  In
comparison, one does not expect this degeneracy in the $\Xi_{cc}$
ground states, since although $s_\de$ is still constrained to equal
1, the (light) third quark is not spatially separated from $\de$, so
that one still expects distinct couplings to the ${\frac{1}{2}}^+$
and ${\frac{3}{2}}^+$ ground states.

Turning now to $L \! > \! 0$ states, the new operators appearing in
the Hamiltonian are pure orbital [{\bf L}$^2$, which is the same for
all states in the $\Sigma^+_g(nL)$ multiplet and therefore provides a
contribution to $M_0$], spin-orbit, and tensor operators.  Both of
the latter operators are considered in Ref.~\cite{Giron:2020fvd} for
$P$-wave $c\bar c q\bar q^\prime$ states.

The spin-orbit operator in this model appears as
\begin{eqnarray}
\Delta H_{LS} & = & V_{LS} \; {\bf L} \cdot \! \left( {\bf s}_\de +
{\bf s}_\bde \right) = V_{LS} \, {\bf L} \cdot {\bf S}\, ,
\label{eq:SpinOrbit}
\end{eqnarray}
where $S$ is the total spin carried by the quarks [the state
subscripts in Eqs.~(\ref{eq:Swavediquark}), or 1 for $Z^{(\prime)}$],
which trivially gives the matrix elements
\begin{equation}
\Delta M_{LS} = \frac{V_{LS}}{2} [J(J+1)-L(L+1)-S(S+1)] \, .
\end{equation}
Note that according to Eq.~(\ref{eq:SpinOrbit}) the model treats all
four quarks on the same footing, each interacting with the same total
{\bf L} operator since the individual diquarks are assumed to have no
internal excitation.  Thus, only one separation coordinate
(${\bf r}_\de - {\bf r}_{\bde}$) and only one orbital angular
momentum operator {\bf L} is relevant.\footnote{Alternate 
$c\bar c c\bar c$ tetraquark models ({\it e.g.},
Refs.~\cite{Badalian:1985es,Liu:2020eha})  have been presented in
which all four quarks and their 3 relative separations are
significant for a full description of the state.}

The final operator in the model for $L \! > \! 0$ states is the
tensor coupling $S_{12}$ between the $\de$-$\bde$ pair, defined by
\begin{equation}
\label{eq:TensorHam}
\Delta H_T = V_T \, S_{12} \, ,
\end{equation}
where
\begin{equation}
\label{eq:Tensor}
S_{12} \equiv 3 \, {\bm \sigma}_1 \! \cdot {\bm r} \, {\bm \sigma}_2
\! \cdot {\bm r} / r^2 - {\bm \sigma}_1 \! \cdot {\bm \sigma}_2 \, .
\end{equation}
${\bm \sigma}$ here and below denotes twice the canonically
normalized spin operator of the full entity coupling to the tensor
force.  In the study of $P$-wave $c\bar c q\bar q^\prime$ states in
Ref.~\cite{Giron:2020fvd} the tensor operator is assumed to originate
as an analogue to the corresponding operator in nucleon-nucleon
interactions arising from pion exchange, and therefore ${\bm \sigma}$
couples only to the light quarks within $\de$ and $\bde$, just as for
the spin-spin $V_0$ operator in Eq.~(\ref{eq:Ham}).  The assumption
of coupling only to the light quarks rather than to the full
$\de,\bde$ as units is viable in the dynamical diquark model because
again, the diquarks are {\em not\/} treated as completely pointlike.
Nevertheless, the alternative hypothesis of coupling the
isospin-dependent spin-spin and tensor operators to $\de,\bde$ as
units was also studied in Refs.~\cite{Giron:2019cfc,Giron:2020fvd}
and found to be incompatible with known phenomenology [{\it e.g.},
in predicting a degenerate $I \! = \! 1$ partner to the $X(3872)$,
which is known not to exist].

In the all-heavy case one not only expects that $\de,\bde$ are more
compact than in the $Q\overline Q q\bar q^\prime$ case, but also
notes that the privileged position of light quarks with respect to
isospin no longer occurs.  In this case, the spin operators
${\bm \sigma}$  in the tensor operator of Eq.~(\ref{eq:Tensor}) refer
to the full $QQ$ or $\overline Q \, \overline Q$ diquark spins.  The
matrix elements in that case are computed in Appendix~A of
Ref.~\cite{Giron:2020fvd}:
\begin{widetext}
\begin{eqnarray}
\label{eq:S12general}
\left< L',S',J
\right| S_{12} \left| L,S,J \right> & = & (-1)^{S+J} \!
\sqrt{30[L][L'][S][S']} \left\{ \begin{array}{ccc} J & S' & L' \\ 2 &
L & S \end{array} \right\} \! \left( \! \begin{array}{ccc} L' & 2 & L
\\ 0 & 0 & 0 \end{array} \! \right) \! \left\{ \! \begin{array}{ccc}
s_\de & s_\bde & S \\ s_{{\de}^\prime} & s_{{\bde}^\prime} & S' \\ 1
& 1 & 2 \end{array} \! \right\} \! \left< s_{{\de}^\prime} ||
\bm{\sigma}_1 || s_{\de} \right> \left< s_{{\bde}^\prime} ||
\bm{\sigma}_2 || s_\bde \right> , \nonumber \\
\end{eqnarray}
\end{widetext}
where $[j] \! \equiv \! 2j \! + \! 1$. The reduced matrix elements of
the angular momentum generators are given by
\begin{equation}
\label{eq:Jreduced}
\left< j^\prime || \, {\bf j} \, || \, j \right> =
\sqrt{j(2j+1)(j+1)} \, \delta_{j^\prime j} \, . 
\end{equation}
The tensor operator of Eq.~(\ref{eq:Tensor}) does not change
individual diquark spins [as is evident from
Eq.~(\ref{eq:Jreduced})], and vanishes if $s_\de \! = \! 0$ or
$s_\bde \! = \! 0$ [as is evident from the $9j$ symbol in
Eq.~(\ref{eq:S12general})].  It does however allow the total quark
spin $S$ to change, as well as the orbital excitation $L$.

In summary, the full Hamiltonian of the dynamical diquark model for
all-heavy states $Q\overline Q Q \overline Q$ (and with small
modifications, for general all-heavy states $Q_1 \overline Q_2 Q_3
\overline Q_4$) is given by the sum of Eqs.~(\ref{eq:Swavecccc}),
(\ref{eq:SpinOrbit}), and (\ref{eq:TensorHam}):
\begin{eqnarray}
H & = & H_0 + 2 \kappa_{QQ} ({\bf s}_Q \! \cdot \!
{\bf s}_Q + {\bf s}_{\bar Q} \! \cdot \! {\bf s}_{\bar Q}) + V_{LS}
\, {\bf L} \cdot {\bf S} + V_T \,S_{12}^{(\de \bde)} . \nonumber \\
\label{eq:FullHam}
\end{eqnarray}
Only the first two terms are required for $\Sigma^+_g(nS)$ states,
while the latter two terms are needed for $L \! > \! 0$ states.  The
matrix elements ({\it i.e.}, mass eigenvalues) for the 3 $S$-wave
states are degenerate and are given in Eq.~(\ref{eq:SwaveMass}),
while those for the $7$ $P$-wave states are presented in
Table~\ref{tab:MassParams}.  The latter are listed in a particular
order that recognizes another interesting feature of this model:
{\em If $V_{LS} \! \gg \! V_T$, then the $P$-wave states fill an
equal-spaced multiplet.}  Assuming that $V_{LS} \! > \! 0$ (as occurs
in
Ref.~\cite{Giron:2020fvd}) means that the states in
Table~\ref{tab:MassParams} may be expected to appear in order of
increasing mass.  This ordering almost precisely matches the
corresponding (unmixed) numbers in Ref.~\cite{Liu:2020eha}, despite
the fact that the latter calculation includes not only tensor terms,
but also couplings between all of the quarks.\footnote{In their full
calculation, Ref.~\cite{Liu:2020eha} also includes color-sextet
combinations.}

The only $\Sigma^+_g (1P)$ states degenerate in $J^{PC}$ are the
$1^{--}$ pair $X_2^{(1)}$ and $X^{\prime \, (1)}_0$.  In that case,
for $V_T \! \neq \! 0$ the states form a $2 \! \times \! 2$ mass
matrix whose diagonal values are given in Table~\ref{tab:MassParams},
and whose off-diagonal element is
\begin{equation}
\Delta M_{X_2^{(1)} \mbox{-} X^{\prime \, (1)}_0}
= +\frac{8}{\sqrt{5}} V_T \, .
\label{eq:Mix}
\end{equation}
\begin{table}[h]
\caption{Mass eigenvalues of the 7 $\Sigma^+_g(nP)$ states, which
assume the simple forms $M \! = \! M_0 \! + \! \kappa_{QQ} \! + \!
\Delta M_{LS} \! + \! \Delta M_T$.  The two $1^{--}$ states
$X_2^{(1)},X^{\prime \, (0)}_1$ also have an off-diagonal mixing term
given by Eq.~(\ref{eq:Mix}).}
\label{tab:MassParams}
\centering
\setlength{\extrarowheight}{1.2ex}
\begin{tabular}{rcrr}
\hline\hline
State & $J^{PC}$ & $\Delta M_{LS}$ & $\Delta M_T$ \\[0.5ex]
\hline
$X_2^{(1)}$ & $ 1^{--}$ & $-3V_{LS}$ & $-\frac{28}{5} V_T$ \\
$Z^{\prime \, (0)}$ & $ 0^{-+}$ & $-2V_{LS}$ & $-8V_T$ \\
$Z^{\prime \, (1)}$ & $ 1^{-+}$ & $-V_{LS}$ & $+4V_T$ \\
$X_2^{(2)}$ & $ 2^{--}$ & $-V_{LS}$ & $+\frac{28}{5} V_T$ \\
$X_0^{\prime \, (1)}$ & $ 1^{--}$ & $0 V_{LS}$ & $0V_T$ \\
$Z^{\prime \, (2)}$ & $ 2^{-+}$ & $+V_{LS}$ & $-\frac 4 5 V_T$ \\
$X_2^{(3)}$ & $ 3^{--}$ & $+2V_{LS}$ & $-\frac 8 5 V_T$ \\[0.5ex]
\hline\hline
\end{tabular}
\end{table}

\section{Numerical Analysis} \label{sec:Num}

LHCb analyzes the results of their observations~\cite{Aaij:2020fnh}
by providing fits to two model scenarios:
\renewcommand{\labelenumi}{\Roman{enumi}.}
\begin{enumerate}
\item $X(6900)$ has $m \! = \! 6905 \pm \! 11$~MeV and
$\Gamma \! = \! 80 \pm \! 19$~MeV\@.  The second resonance,
hereinafter labeled $X(6500)$, lies at $6490 \pm \!
15$~MeV.\footnote{This value is not stated in
Ref.~\cite{Aaij:2020fnh}, but rather is estimated by us using their
Fig.~3(b).}  The mass splitting between these states is
$\Delta m_{\rm I} \! = \! 415 \! \pm \! 19$~MeV\@.
\item $X(6900)$ has $m \! = \! 6886 \! \pm \! 11$~MeV and
$\Gamma \! = \! 168 \! \pm \! 33$~MeV\@.  The second resonance,
hereinafter labeled $X(6740)$, has $m \! = \! 6741 \! \pm \! 6$~MeV
and $\Gamma \! = \! 288 \! \pm \! 16$~MeV\@.  The mass splitting
between these states is
$\Delta m_{\rm II} \! = \! 145 \! \pm \! 15$~MeV\@. 
\end{enumerate}
We now show that the scenario of Model~II appears to support a much
more favorable interpretation within the dynamical diquark model.

For this analysis we first assume that $X(6900)$ is not a $1S$ state,
because it would then lie 700~MeV above the $J/\psi$-pair threshold,
which would represent an astonishing mass gap for the appearance of
the lowest $c\bar c c\bar c$ resonances.  Similar conclusions appear
in Refs.~\cite{Liu:2020eha,Wang:2020ols,Jin:2020jfc,Yang:2020rih,
Becchi:2020uvq,Lu:2020cns,Chen:2020xwe}.  We discuss the fate of the
$1S$ states in our model later in this section; the subsequent
multiplets in order of increasing mass turn out to be $1P$, $2S$,
$1D$, $2P$, and $2D$, as confirmed below.

The next required input of the analysis is a reliable value of the
internal diquark spin-spin coupling $\kappa_{cc}$ appearing in
Eqs.~(\ref{eq:Swavecccc})--(\ref{eq:SwaveMass}).  The closest
available analogue to $c\bar c c\bar c$ state is found with
$c\bar c s\bar s$ candidates such as $X(4140)$, which have been
analyzed using this model very recently in Ref.~\cite{Giron:2020qpb}.
In that work, $\kappa_{cs}$ is found to be quite large (114.2~MeV)
compared to the fit value for $\kappa_{cq}$ or $\kappa_{bq}$
(17.9--22.5~MeV).  We observed in Ref.~\cite{Giron:2020qpb} that this
pattern is explained by the diquark coupling being strongly dependent
upon the lighter quark flavor ($\kappa_{cs}$ {\it vs.}\
$\kappa_{cq}$) and much less sensitive to the heavy-quark flavor
($\kappa_{cq}$ {\it vs.}\ $\kappa_{bq}$).  We argued that the $s$
quark, being much heavier than $u$ or $d$, has less Fermi motion
within $\de$, permitting $\de$ to be substantially more compact and
thus enhancing the strength of spin couplings within it.  Therefore,
it is reasonable to assume that the ($cc$) diquark has a similarly
large spin-spin coupling (and possibly even larger, if $s$ is
insufficiently heavy to reach the point of flavor independence for
the lighter quark in $\de$).  Hence, for all states in this fit we
take the spin-spin coupling to be
\begin{equation}
\kappa_{cc} = 114.2 \ {\rm MeV} \, .
\label{eq:KappaValue}
\end{equation}
Note from Eq.~(\ref{eq:SwaveMass}) or Table~\ref{tab:MassParams} that
such a large value of $\kappa_{cc}$ leads to the interesting
consequence of predicting $M_0$, and hence the diquark mass $m_\de$,
to be rather smaller than in fits from other works.

We now possess sufficient information to study $S$-wave multiplet
masses, as well as $P$-wave multiplet masses ignoring for the moment
the spin-orbit and tensor terms.  Two natural assignments for
$X(6900)$ may be considered: as a $\Sigma_g^{+}(1P)$ or as a
$\Sigma_g^{+}(2S)$ state.  One then calculates for each case the mass
splittings to lower multiplets, in order to confirm whether one or
both of these assignments matches the mass splittings
$\Delta m_{\rm I}$ and/or $\Delta m_{\rm II}$ between peaks from
LHCb's Model~I or II, respectively.

First we investigate the possibility that $X(6900)$ is a
$\Sigma_g^{+}(1P)$ state.  Since the $J/\psi$ pair has $C \! = \! +$,
Table~\ref{tab:MassParams} suggests that the lightest allowed
candidate (assuming $V_{LS}, V_T \! > \! 0$, as is used below) is
$Z^{\prime \, (0)} (0^{-+})$.  To be quantitative, we adopt the
numerical results obtained from the $P$-wave $c\bar c q\bar q$ states
in Ref.~\cite{Giron:2020fvd}.  Specifically, we use values obtained
from Cases~3 and 5 of Ref.~\cite{Giron:2020fvd} for $V_{LS}$ and
$V_T$, which are
\begin{equation}
V_{LS}=42.9\;\mathrm{MeV}, \; V_T=5.5\;\mathrm{MeV},
\label{eq:NumFit1}
\end{equation}
and
\begin{equation}
V_{LS}=49.0\;\mathrm{MeV}, \; V_T=3.8\;\mathrm{MeV},
\label{eq:NumFit2}
\end{equation}
respectively.  These cases were deemed in Ref.~\cite{Giron:2020fvd}
to be the ones most likely to accurately represent the true $P$-wave
$c\bar c q\bar q^\prime$ spectrum.  Their application to the
$c\bar c c\bar c$ system deserves some discussion.  The spin-orbit
term in this model connects two separated heavy diquarks in either
case [($cq$) or ($cc$)], and therefore we assume the size of the
coupling $V_{LS}$ to depend upon the source only through its spin and
not its flavor content, so long as the diquarks are heavy.  The
tensor term, on the other hand, is an entirely different matter.  In
Ref.~\cite{Giron:2020fvd} the tensor operator was chosen to couple
only to light-quark spins [see the discussion below
Eq.~(\ref{eq:Tensor})], while the $c\bar c q\bar q^\prime$ analogue
to the form of Eq.~(\ref{eq:TensorHam}) used here for
$c\bar c c\bar c$ was found to be phenomenologically irrelevant.  We
therefore take as our final assumption that $V_T$ for
$c\bar c c\bar c$ is numerically no larger than the $V_T$ values
obtained from $c\bar c q\bar q^\prime$.

Using the values for $\kappa_{cc}, V_{LS}, V_T$ in
Eqs.~(\ref{eq:KappaValue})--(\ref{eq:NumFit2}), one then needs only
the mass expressions in Table~\ref{tab:MassParams} and
Eqs.~(\ref{eq:SwaveMass}) and (\ref{eq:Mix}).  Fixing the
$Z^{\prime \, (0)}$ mass eigenvalue to the (Model~I) $X(6900)$ mass,
we implement the Schr\"{o}dinger equation-solving numerical
techniques applied to lattice-calculated potentials, as described in
Ref.~\cite{Giron:2019bcs}.  We thus obtain
\begin{equation}
M_0(1P) = 6931.3 \ {\rm MeV \ and} \ 6954.0 \ {\rm MeV} \, ,
\end{equation}
using the inputs of Eqs.~(\ref{eq:NumFit1}) and (\ref{eq:NumFit2}),
respectively.\footnote{The variation of these particular eigenvalues
with the lattice potentials obtained in Refs.~\cite{Juge:1997nc,
Juge:1999ie,Juge:2002br,Morningstar:2019,Capitani:2018rox} amounts to
only about 0.07~MeV\@.  The specific values presented here use
Ref.~\cite{Morningstar:2019}.}  Further computing $M_0(1S)$ and
$M_0(2S)$ in the same calculation, we obtain the $M_0$ mass
differences
\begin{eqnarray}
\Delta m_{1P - 1S} & = & +343.3 \ {\rm MeV} \, , \nonumber \\
\Delta m_{1P - 2S} & = & -156.9 \ {\rm MeV} \, ,
\label{eq:1PFit}
\end{eqnarray}
using Eqs.~(\ref{eq:NumFit1}).  The corresponding values obtained
using Eqs.~(\ref{eq:NumFit2}) are hardly changed, being +343.2~MeV
and $-156.7$~MeV, respectively.  In comparison with the LHCb results,
the first of Eqs.~(\ref{eq:1PFit}) is too small to match Model~I
({\it i.e.}, $\Delta m_{1P - 1S} \! < \! \Delta m_{\rm I}$),
especially since $M_0(1P)$ lies rather higher than the
$Z^{\prime \, (0)}$ mass we fix to $X(6900)$, while the second has
the right magnitude but the wrong sign to match Model~II ({\it i.e.},
$\Delta m_{\rm II} \! \approx \! -\Delta m_{1P - 2S}$), since we
predict that $2S$ states lie above $1P$ states.  We therefore
conclude that the assignment of $X(6900)$ as a $\Sigma_g^+(1P)$ state
is heavily disfavored in the dynamical diquark model. 

We therefore turn to the alternate possibility that $X(6900)$ is one
of the states in the multiplet $\Sigma_g^{+}(2S)$ (which again, are
degenerate in this model).  Then using Eqs.~(\ref{eq:SwaveMass}),
(\ref{eq:KappaValue}), and the Model-II mass value, we obtain
\begin{equation}
M_0(2S) = 6771.8 \ {\rm MeV} \, .
\label{eq:True6900}
\end{equation}
Once again implementing the techniques developed in
Ref.~\cite{Giron:2019bcs}, we calculate the $M_0$ mass differences
\begin{eqnarray}
\Delta m_{2S - 1P} & = & 160.4 \  {\rm MeV} \, , \nonumber \\
\Delta m_{2S - 1S} & = & 505.7 \ {\rm MeV} \, .
\end{eqnarray}
In this case we observe that the latter mass splitting is too large
to agree with Model~I ({\it i.e.}, $\Delta m_{2S - 1S} \! >
\Delta m_{\rm I}$), but the former agrees very well with Model~II
({\it i.e.}, $\Delta m_{2S - 1P} \! \approx \! \Delta m_{\rm II}$).
Therefore, assuming that LHCb's Model~II is confirmed to be the
correct interpretation of the data, we find that $X(6900)$ is favored
in the dynamical diquark model to be a $\Sigma_g^{+}(2S)$ state and
$X(6740)$ a $\Sigma_g^{+}(1P)$ state. 

Concluding from these calculations that $X(6900)$ is indeed a
$\Sigma_g^{+}(2S)$ state with $M_0(2S)$ given by
Eq.~(\ref{eq:True6900}), the corresponding diquark masses are
computed to be
\begin{equation}\label{diquark_mass_2S}
m_\de=m_\bde=3126.4 \mbox{-} 3146.4\;\mathrm{MeV} ,
\end{equation}
which is only slightly larger than $m_{J/\psi}$.  Using this value of
$m_\de$, we further obtain
\begin{eqnarray}
M_0(1S) & = & 6264.0 \mbox{-} 6266.1\;\mathrm{MeV}, \nonumber \\
M_0(1P) & = & 6611.4\;\mathrm{MeV}, \nonumber \\
M_0(1D) & = & 6860.5 \mbox{-} 6862.4\;\mathrm{MeV}, \nonumber \\
M_0(2P) & = & 7010.8 \mbox{-} 7013.0 \; \mathrm{MeV} .
\end{eqnarray}
The variation here arises from using the differing lattice results of
Refs.~\cite{Juge:1997nc,Juge:1999ie,Juge:2002br,Morningstar:2019,
Capitani:2018rox}.  The prediction for $M_0(1S)$ deserves special
discussion, because the expected spatial size of a $1S$ state
according to this model is calculated to be $\langle r \rangle \!
\approx \! 0.3$~fm, the same magnitude as (or even smaller than)
$J/\psi$ states.  In this scenario all 4 of the quarks have
comparable spatial separation, a configuration that runs afoul of the
original separated-diquark motivation of the dynamical diquark model.
At present, the LHCb data in the $\sim$~6300~MeV mass region is not
yet sufficiently resolved to discern particular structures, so it
will be interesting to see how well the model works even in
situations for which it is expected to fail.  

Having identified $X(6900)$ with one of the (degenerate)
$\Sigma_g^{+}(2S)$ states, we use the values of $V_{LS}$ and $V_T$
given by Eqs.~(\ref{eq:NumFit1}) and (\ref{eq:NumFit2}) and the
expressions in Table~\ref{tab:MassParams} and Eq.~(\ref{eq:Mix}) to
compute the full $\Sigma^+_g(1P)$ spectrum.  The results are
presented in  Table~\ref{tab:BestFit}.  One notes that the variation
in mass for any given state between the two fits [excepting
$X_2^{(2)}(2^{--})$] is $\alt \! 13$~MeV, and that the ordering of
the states in mass is nearly identical to the one expected
parametrically from the equal-spacing rule identified in
Table~\ref{tab:MassParams}, even though the equal-spacing itself is
numerically not so well supported.  Since the values of $V_T$ in
Eqs.~(\ref{eq:NumFit1})--(\ref{eq:NumFit2}) are based upon a naive
assumption, the equal-spacing rule might turn out to be much better
in practice if the actual $V_T$ value is smaller.  

An interesting feature of LHCb Model~II is the enormous width
$\Gamma \! = \! 288$~MeV given for $X(6740)$ (twice the width of
$\rho$, for example).  From Table~\ref{tab:BestFit} we note that all
$P$-wave states that could decay to a $J/\psi$ pair ($C \! = \! +$)
have masses consistent with appearing within this wide peak, meaning
that the broad $X(6740)$ peak could easily turn out to be a
superposition of several narrower $1P$-state peaks.
\begin{table}[h]
\caption{Mass eigenvalues (in MeV) of the 7 $\Sigma^+_g(1P)$ states,
using the expressions given in Table~\ref{tab:MassParams} and
Eq.~(\ref{eq:Mix}).  $M_0(1P)$ is obtained from the same numerical
fit identifying $X(6900)$ as a $\Sigma^+_g(2S)$ state (specifically,
using the lattice simulation of Ref.~\cite{Morningstar:2019}),
$\kappa_{cc}$ is given in Eq.~(\ref{eq:KappaValue}), and the columns
represents two different choices for $V_{LS}$ and $V_T$ values.}
\label{tab:BestFit}
\centering
\setlength{\extrarowheight}{1.2ex}
\begin{tabular}{rcrr}
\hline\hline
State & $J^{PC}$ & Eq.~(\ref{eq:NumFit1}) & Eq.~(\ref{eq:NumFit2}) \\
[0.5ex]
\hline
$X_2^{(1)}$ & $ 1^{--}$ & $6563.70$ & $6556.22$ \\
$Z^{\prime \, (0)}$ & $ 0^{-+}$ & $6595.79$ & $6597.19$ \\
$Z^{\prime \, (1)}$ & $ 1^{-+}$ & $6704.69$ & $6691.79$ \\
$X_2^{(2)}$ & $ 2^{--}$ & $6713.49$ & $6687.87$ \\
$X_0^{\prime \, (1)}$ & $ 1^{--}$ & $6727.98$ & $6726.68$ \\
$Z^{\prime \, (2)}$ & $ 2^{-+}$ & $6764.09$ & $6771.55$ \\
$X_2^{(3)}$ & $ 3^{--}$ & $6802.59$ & $6817.51$ \\[0.5ex]
\hline\hline
\end{tabular}
\end{table}

Finally, a notable enhancement in the LHCb data appears slightly
above $7200$~MeV\@.  This value coincides with the
$\Xi_{cc}$-$\overline{\Xi}_{cc}$ threshold 7242.4~MeV, at which
sufficient energy becomes available to create the lightest hadronic
state containing both $c\bar c c\bar c$ and an additional light
$q\bar q$ valence pair, namely, the baryon pair
$(ccq)(\bar c \bar c \bar q)$.  Above this threshold one expects no
further narrow resonances decaying dominantly to $J/\psi$ pairs,
since new open-flavor decay channels become kinematically available.
This prediction is particularly easy to see in the dynamical diquark
model; it is the point at which the gluon flux tube connecting the
$\de$-$\bde$ pair gains enough energy to fragment through $q\bar q$
pair creation, and was anticipated in Ref.~\cite{Brodsky:2014xia} for
$c\bar c q\bar q$ states to occur at the
$\Lambda_c^+$-$\bar{\Lambda}_c^-$ threshold.  Interestingly, we find
the $2D$ states to have a common multiplet mass of
\begin{equation}
M_0(2D)=7213.3 \mbox{-} 7216.7\;\mathrm{MeV},
\end{equation}
meaning that the enhancement in the data above 7200~MeV may be a
combination of some $2P$ and/or $2D$ $c\bar c c\bar c$ states [not
forgetting the large mass offset due to $\kappa_{cc}$ from
Eqs.~(\ref{eq:FullHam}) and (\ref{eq:KappaValue})] with threshold
effects in the form of rescattering of
$\Xi_{cc}$-$\overline{\Xi}_{cc}$ pairs to $J/\psi$ pairs.  In
addition, the $c\bar c c\bar c$ states in higher BO multiplets than
$\Sigma^+_g$ ({\it i.e.}, analogues to hybrid mesons) would also
occur at or above the $\Xi_{cc}$-$\overline{\Xi}_{cc}$ threshold.

\section{Conclusions} \label{sec:Concl}

The recent LHCb discovery of resonance-like structures in the
$J/\psi$-pair spectrum opens a whole new arena for hadronic
spectroscopy. The $X(6900)$ represents the first clear candidate for
a multiquark exotic hadron that contains only heavy valence quarks.
This paper and multiple prior works referenced here suggest that
numerous other such states, carrying a variety of quantum numbers,
await discovery as experimental observations are refined.
Furthermore, the all-heavy sector is particularly interesting from a
theoretical point of view, since the molecular binding paradigm
popular for light-flavor containing multiquark states like $X(3872)$
is much less viable (particularly for states that lie so far above
the $J/\psi$-pair threshold), leaving a diquark-antidiquark binding
structure as the leading candidate.

This paper has explored the basic spectroscopic properties of the
all-heavy 4-quark states $Q_1 \overline Q_2 Q_3 \overline Q_4$ in the
dynamical diquark model. Its defining features for this system are
(1) the dominance of the color-triplet binding between $\de \! \equiv
\! Q_1 Q_3$ and between $\bde \! \equiv \! \overline Q_2 \overline
Q_4$, which for the identical-quark cases $c\bar c c\bar c$ or
$b\bar b b\bar b$ leads to the absence of $1^{++}$ $S$-wave states;
(2) the dominance of spin-spin couplings within $\de$ and within
$\bde$, but not between quarks and antiquarks, which leads to the
degeneracy of all 3 states in each $Q\overline Q Q\overline Q$
$S$-wave multiplet; and (3) a spin-orbit coupling for $L \! > \! 0$
that couples to all quarks with the same strength. If the strength
of the tensor coupling is substantially smaller than the spin-orbit
coupling, then the 7 states of the $P$-wave $Q\overline Q
Q\overline Q$ multiplet exhibit a remarkable equal-spacing spectrum.
These features clearly provide simple and immediate tests of various
aspects of the model.

We have also produced numerical predictions of the full spectrum for
the $1S$, $1P$, and $2S$ multiplets, and multiplet-averaged masses
for $1D$, $2P$, and $2D$, using lattice-calculated confining
potentials, the spin-spin coupling obtained from $c\bar c s\bar s$
candidate states, and the spin-orbit and tensor couplings obtained
from $P$-wave $c\bar c q\bar q^\prime$ states, all using this model.
In attempting different assignments for the $X(6900)$, we find that
the only one compatible with the model is to identify $X(6900)$ with
a state or states within the $2S$ multiplet, and the lower structure
at about 6740~MeV from LHCb's ``Model II'' being some combination of
the $C \! = \! +$ states within the $1P$ multiplet. Evidence for the
$1S$ multiplet is obscure, possibly because it is predicted to occur
at masses at which the $\de$-$\bde$ structure is no longer viable,
since all interquark distances become comparable not far above the
$J/\psi$-pair threshold, while $1D$ states could easily be obscured
by the large $X(6900)$ peak, and some $2P$ and $2D$ states are
predicted to lie at or above the $\Xi_{cc}$-$\bar \Xi_{cc}$ threshold
(which coincides with a structure in the LHCb results), at which
point the $c\bar c c\bar c$ states are expected to become much wider.

The resolution of the newly observed $J/\psi$-pair structures
(possibly into several peaks) and the measurement of specific
$J^{PC}$ quantum numbers will contribute immeasurably to an
understanding of the structure of these states. Future studies of
other charmonium-pair structures (including $\chi_c$, $h_c$, and
$\eta_c$) will be no less valuable in this regard.

\begin{acknowledgments} This work was supported by the National
Science Foundation (NSF) under Grant No.\ PHY-1803912.
\end{acknowledgments}

\bibliographystyle{apsrev4-1}
\bibliography{diquark}
\end{document}